\documentclass{article}
\usepackage{arxiv}
\usepackage[utf8]{inputenc} % allow utf-8 input
\usepackage[T1]{fontenc}    % use 8-bit T1 fonts
\usepackage{hyperref}
\hypersetup{
    colorlinks=true,
    linkcolor=blue,
    filecolor=magenta,      
    urlcolor=cyan,
    }
\usepackage{upgreek}% hyperlinks
\usepackage{url}            % simple URL typesetting
\usepackage{booktabs}       % professional-quality tables
\usepackage{amsfonts}       % blackboard math symbols
\usepackage{nicefrac}       % compact symbols for 1/2, etc.
\usepackage{microtype}      % microtypography
\usepackage{lipsum}
\usepackage{float}
\usepackage{graphicx}
\usepackage{multirow}
\usepackage{array}
\usepackage{comment}
\usepackage{longtable}
\usepackage{ragged2e}
\usepackage{subfig}
\usepackage{amsmath}
\usepackage{longtable}
\usepackage{ amssymb }
\usepackage{amsmath}
\title{Introduction to Medical Imaging Informatics}

\author{
  Md. Zihad Bin Jahangir\\
  Research and Development Department, Pioneer Alpha\\
Dhaka, Bangladesh\\
\texttt{zihad.bscincse@gmail.com}\\
   \And
    Ruksat Hossain   \\
  Research and Development Department, Pioneer Alpha,\\
Dhaka, Bangladesh\\
\texttt{ruksatboni@gmail.com}
   \And
      Riadul Islam \\
  Research and Development Department, Pioneer Alpha,\\
Dhaka, Bangladesh\\
\texttt{riyadcseiiuc72@gmail.com}
   \And
MD Abdullah Al Nasim\\
 Research and Development Department, Pioneer Alpha,\\
Dhaka, Bangladesh\\
\texttt{nasim.abdullah@ieee.org} \\
   \And 
Md. Mahim Anjum Haque\\
Research and Development Department, Pioneer Alpha,\\
Dhaka, Bangladesh\\
\texttt{mahim@vt.edu}\\
   \And
 Md Jahangir Alam \\
  Department of Computer Science\\
  University of Alabama at Birmingham\\
Alabama, USA\\
  \texttt{malam@uab.edu} \\
   \And
   Sajedul Talukder\\
   Department of Computer Science\\
   University of Alabama at Birmingham\\
  Alabama, USA\\
  \texttt{stalukder@uab.edu}}
\begin{document}
\maketitle

\begin{abstract}
Medical imaging informatics is a rapidly growing field that combines the principles of medical imaging and informatics to improve the acquisition, management, and interpretation of medical images. This chapter introduces the basic concepts of medical imaging informatics, including image processing, feature engineering, and machine learning. It also discusses the recent advancements in computer vision and deep learning technologies and how they are used to develop new quantitative image markers and prediction models for disease detection, diagnosis, and prognosis prediction. By covering the basic knowledge of medical imaging informatics, this chapter provides a foundation for understanding the role of informatics in medicine and its potential impact on patient care. 

\end{abstract}

\section{Introduction}

Medical Imaging Informatics is an expeditiously advancing area enclosed by the continuous process of analyzing and improving supervision of patient data, clinical knowledge and any other information relevant to community health through the application of methods and bridging between imaging and other medical disciplines \cite{1}. Starting from an elaboration of medical imaging informatics, this chapter includes basic knowledge of image processing, idea and illustration of feature engineering and machine learning incorporating state-of-the-art advancements in computer vision and different aspects of deep learning technologies to develop new quantitative image markers or prediction models for disease detection and prognosis prediction. To make the chapter comprehensive, we have tried to demonstrate recent advancements, and current challenges leading to future work. The field of medical imaging informatics involves using information and communication technologies (ICT) in medical imaging to provide healthcare services. This field has grown significantly over the past three decades to encompass a range of multi-disciplinary medical imaging services, from routine clinical practices to the study of advanced human physiology and pathophysiology. The Society for Imaging Informatics in Medicine (SIIM) originally defined this field as~\cite{2,3}:

\textit{"Imaging informatics touches every aspect of the imaging chain from image creation and acquisition to image interpretation, reporting, and communications. The field serves as the integrative catalyst for these processes and forms a bridge with imaging and other medical disciplines."}

Medical imaging informatics combines the principles of medical imaging and informatics to improve the acquisition, management, and interpretation of medical images. It involves using advanced computational techniques and software to analyze and interpret medical images, such as X-rays, CT scans, and MRI scans.

The purpose of medical imaging informatics, as defined by the Society for Imaging Informatics in Medicine (SIIM), is to enhance the effectiveness, accuracy, and reliability of medical imaging services within the healthcare system~\cite{3}. This includes using and exchanging medical images within complex healthcare systems~\cite{3} and incorporating technological advancements such as big data imaging, electronic health records analytics, and workflow optimization. These innovations are driving the development of precision medicine and shaping the future of medical imaging informatics~\cite{4,5,6}.
In this field, image processing techniques are used to enhance the visibility of specific structures in medical images, extract features from images~\cite{9}, and analyze the images to identify abnormalities or other characteristics of interest~\cite{10,11}. In medical imaging informatics, we may need feature engineering involves extracting and selecting relevant features from medical images to train a machine-learning model to recognize patterns or features of interest in that images~\cite{12}.

Machine learning models are also used in medical imaging informatics to develop predictive models for disease detection, diagnosis, and prognosis prediction~\cite{13,14,biswas2022mri}. These models are trained on large datasets of medical images and associated clinical data. They can be used to classify images automatically or predict outcomes based on the features extracted from the images~\cite{13,14,15}.
Recent advancements in computer vision and deep learning technologies have led to significant progress in medical imaging informatics. Deep learning algorithms, in particular, have shown great promise in developing new quantitative image markers and prediction models for various medical applications.

\section{Literature review}

The interpretation of medical images can be a complex and time-consuming task, and there is a need for more efficient and accurate methods for analyzing and interpreting these images \cite{biswas2022mri}.
To address this need, researchers and practitioners have developed a range of techniques and technologies for medical imaging informatics. These techniques include image processing algorithms to enhance the visibility of certain structures in medical images, software tools for measuring and analyzing images, and systems for storing and sharing medical images across different healthcare organizations.
In recent years, significant advancements in computer vision and deep learning technologies have opened up new possibilities for medical imaging informatics. These technologies have been used to develop new quantitative image markers and prediction models for various medical applications, including disease detection, diagnosis, and prognosis prediction.
Overall, medical imaging informatics plays a crucial role in medicine, helping healthcare providers more accurately and efficiently analyze and interpret medical images and make more informed patient care decisions.

\section{Medical imaging informatics}

Medical Imaging Informatics is a revolutionary subset of medical informatics that encompasses image coding, image processing, image distribution connection to image acquisition devices with analogue or digital output and communicating information (data) that is pivotal to the provision and delivering appropriate patient care critical for health and well-being. The revolution of medical imaging and biomedical informatics has changed the confined way of research and the nature of medicine. At all major levels of health care and in a variety of medical settings, medical imaging plays a pivotal role. Technologies like X-rays, mammography, computed tomography (CT scans), ultrasonography and other clinical specialities are the edging of medical imaging. Recent advances in medical imaging technology, e.g. Picture Archive and Communication systems (PACS), image-guided surgery and therapy, computer-aided diagnosis (CAD), and electronic Patient Record (ePR) with image distribution have propelled imaging informatics as a discipline to manage and synthesize knowledge from medical images for effective and efficient patient care as well as outcomes~\cite{2}.
On the contrary, biomedical informatics incorporates a wide range of domain-specific methodologies. For the betterment of human health, biomedical informatics is the integrative field that studies the effective uses of biomedical data for scientific inquiry, problem-solving and decision-making. Biomedical informatics integrates computer applications ranging from the processing of very low-level narrations to extremely high-level ones, which are completely and systematically different~\cite{3}.

\begin{figure}
    \centering
    \includegraphics[width=.8 \textwidth]{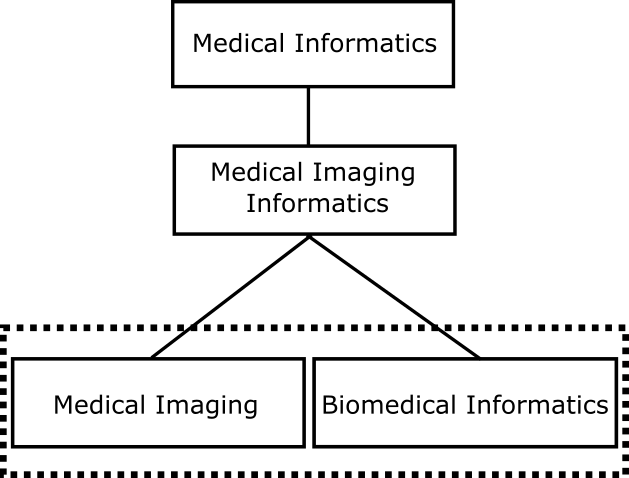}
    
    \caption{Medical Imaging Informatics }
    \label{fig11}
\end{figure}

Medical imaging informatics is a field that involves the use of computer science, information science, and engineering to manage, analyze, and interpret medical images and data. This includes developing and implementing systems and software to store, retrieve, and analyze medical images and integrating these images with electronic medical records and other health information systems.
Artificial intelligence and machine learning are increasingly used in medical image informatics to improve the accuracy, efficiency, and effectiveness of medical imaging \cite{hossainbrain}.

\begin{table*}[h]
  \caption{SOME CHARACTERISTICS OF MEDICAL IMAGING MODALITIES}
    \centering
    \begin{tabular}{|c|p{6cm}|p{2cm}|p{2cm}|}
    \hline
        \textbf{Name} & \textbf{Technology} & \textbf{Anatomies}& \textbf{Dimension} \\
        \hline
         X-ray & 
         X-ray imaging uses a small amount of ionizing radiation to create images of the inside of the body, which can be used to visualize bones and some soft tissues~\cite{7} &
         Most organ &
         2D, 2D+t \\
         \hline
         CT & 
         CT scans use a series of X-rays to create detailed, cross-sectional images of the body. CT scans visualize organs, bones, and other tissues in great detail~\cite{8}. &
         Most organ &
         2D, 3D, 4D \\
         \hline
         Ultrasound & 
         Uses high-frequency sound waves to create images of the inside of the body. Ultrasound is often used to visualize the abdomen, pelvis, and other internal organs~\cite{7}. &
         Most organ &
         2D,2D+t, 3D, 4D \\
         \hline
         MRI & 
         MRI uses a powerful magnetic field and radio waves to create detailed images of the body's tissues and organs. &
         Most organ &
         3D, 4D \\
         \hline
         Nuclear & 
         Utilize external detectors or gamma cameras to detect the emission of gamma rays from radioisotopes that have been ingested~\cite{7}. &
         All organs with radioactive tracer uptake &
         2D, 3D, 4D \\
         \hline
         Microscopy & 
         Typically use an illumination source and lenses to magnify specimens before capturing an image~\cite{7}. &
         Primarily biopsies and surgical specimens &
         2D, 3D, 4D \\
         \hline
    \end{tabular}
    \label{tab:my_label}
\end{table*}

Here are some of the most popular applications for AI in medical image informatics.

\begin{enumerate}
  \item Image analysis and interpretation: Using AI in image analysis and interpretation can improve the speed and accuracy of diagnosis and help healthcare providers make more informed decisions about treatment.
  \item Computer-aided diagnosis: AI can assist radiologists and other healthcare providers in interpreting medical images and identifying abnormalities, which can help improve the accuracy and efficiency of diagnosis.
  \item Image-guided surgery: AI can be used to help surgeons navigate during procedures by providing real-time image guidance and information about the location and orientation of surgical instruments.
  \item Predictive analytics: AI can analyze large amounts of data from medical images and other sources to make predictions about patient outcomes or identify potential risk factors for specific conditions.
\end{enumerate}

\subsection{Types of medical imaging modalities}

There are several different types of medical imaging modalities, each of which uses different methods and technologies to produce images of the body. Some of the most common medical imaging modalities include:

\begin{enumerate}
  \item X-ray: X-ray imaging uses a small amount of ionizing radiation to produce images of the body's internal structures. X-rays are commonly used to visualize bones and can be used to diagnose fractures, osteoporosis, and other bone conditions.
  \item CT (computed tomography): CT uses X-rays and a computer to produce detailed images of the body's internal structures. CT scans often diagnose cancer, cardiovascular disease, and other conditions.
  \item MRI (magnetic resonance imaging): MRI uses a strong magnetic field and radio waves to produce detailed images of the body's soft tissues, such as muscles, tendons, and organs. MRI is often used to diagnose brain and spinal cord injuries and conditions such as multiple sclerosis and cancer.
  \item Ultrasound: Ultrasound uses high-frequency sound waves to produce images of the body's internal structures. Ultrasound is often used to visualize the fetus during pregnancy and the heart, blood vessels, and other organs.
  \item PET (positron emission tomography): PET uses small amounts of radioactive material to produce images of the body's metabolic activity. PET scans are often used to diagnose cancer and other conditions.
\end{enumerate}

\subsection{Image storage and retrieval}

Image storage and retrieval is an essential aspect of medical imaging informatics, as it involves managing and organizing medical images in a way that allows them to be easily accessed by authorized users.
Medical images can be stored electronically in various formats, including DICOM (Digital Imaging and Communications in Medicine), a standard format for storing and transmitting medical images. These images can be stored on a central server or in the cloud and accessed by authorized users, such as doctors and technologists, from any location. There are several benefits to the electronic storage of medical images: Reduced physical storage space, Faster access to images, Improved organization, and Enhanced security. 

\subsection{Image analysis and interpretation}

Image analysis and interpretation using algorithms and software to analyze and interpret medical images to extract meaningful information and insights. This is an important aspect of medical imaging informatics, as it can help doctors and other healthcare professionals more accurately diagnose and treat medical conditions. There are several ways in which image analysis and interpretation can be used in medical imaging:
Automated image analysis: Algorithms and software can automatically analyze medical images and identify specific features or abnormalities. For example, an algorithm might be trained to detect tumors in CT scans. Computer-aided diagnosis: Image analysis and interpretation can assist doctors in diagnosing by providing additional information and insights that might not be apparent from the raw images alone. For example, an algorithm might be used to identify patterns in an MRI scan that suggest the presence of a particular condition. Artificial intelligence and machine learning: Machine learning algorithms can improve image analysis and interpretation by learning from large datasets of labeled images. This can improve the accuracy and efficiency of the analysis process.

\section{Image Processing}
The field of medical imaging (MI), and image processing (IP) have been emerged as one of the fastest-developing research areas in computer vision. A large array of techniques are employed in the subsection of digital signal processing known as “image processing” to improve or edit digital pictures in order to prepare them more useful for a variety of reasons. So the study of computer vision focuses on how computers can “understand” photos, movies, or 3D volumes by extracting the needed elements and properties from the images using a variety of algorithms and methodologies [4]. It serves as the foundation for the model training. Various image editing techniques, such as super-resolution, denoising, dehazing, deraining, and deblurring, are referred to as image processing. One must first understand the concept of a picture before learning how it is processed. Image processing is a technique for applying certain operations to an image in order to produce an improved image or to pull out any meaningful information from that. It is a technique of processing where an image serves as the input, and the output can either be another image or attributes or properties associated with the input image. It is just one of the advanced technology that is now developing swiftly and is a major area of study for both engineers and computer scientists. The terms analogue and digital image processing refer to two different categories of image processing. For tangible copies like prints and pictures, analogue image processing can be employed. When interpreting images utilizing these visual methods, image analysts employ a variety of interpretive basics. Through the use of computers, digital image processing techniques enable picture alteration. As image processing is a fundamental part of model training, it needs to remove image noise for better feeding into the model. In the memoir, the usefulness of computer vision as a tool has been quite limited, since most prospective applications had to wait to develop affordable memory technology and realize suitable perfection ratios in terms of processing power~\cite{5}. Digital image processing (DIP) consists of 11 core phases, each of which may include further steps. The following is a description of the fundamental processes in digital image processing.
To process digital images, one must first complete these basic stages. Getting a picture that is already digitally stored might suffice as an easy method of image acquisition. Pre-processing, involving scaling, etc., usually occurs at the imagery collection stage. One of the easiest and most attractive aspects of digital image processing is picture enhancement. The basic concept underlying enhancement techniques are to either emphasize certain elements of interest within an image or reveal concealed info. Shifting the luminance, for instance.
The process of photo restoration involves improving the appearance of an image. Picture restoration is objective, in contrast to augmentation, which is intuitive because restoration procedures often are based on mathematical or probabilistic models of image deterioration. The enormous expansion in the use of digital photographs on the Internet has led to a growing relevance for the field of color image processing. This might involve, among other things, digital color modeling, and processing. Wavelets serve as the foundation for expressing pictures at different resolution levels. Splitting of pictures into progressively smaller areas for data compression and pyramidal depiction. Compression techniques involve ways to lower the amount of space needed to store or transmit a picture. It is essential to compress data, particularly for web-based applications. The field of morphological processing focuses on methods for separating image elements that help represent and describe the shape. Segmentation techniques separate a picture into its elements or objects. Generally speaking, another of the most challenging jobs in digital image processing is independent fragmentation. Imaging issues that need individual object identification can be successfully solved with the help of a tough segmentation approach.

The recently advanced image processing technique is named MAXIM [6] which has been formed from multi-axis MLP. Due to the absence of focus on the immediate area, it is employed for low-level vision to facilitate improved pictures. This method operates using UNet-shaped hierarchical composition and focuses on long-distance interactions. It takes on two MLP-based building blocks: (i) Multi-axis gates and (ii) a Cross-gating block. We could observe the architecture of Maxim in Figure 2 and it depended on a half-cast design for each block (Figure 2b) and also puts in an encoder, decoder, and bottlenecks with residual channel attention block (RCAB) with filtering of skip connections. In Figure (2a), It has the backbone of MAXIM with a cross-gating block on Convolution (figure 2c). This proposed model acquires performance of the highest calibre of more than ten benchmark datasets for image processing tasks, adding denoising, de-blurring, deraining, and enhancement whilst using less or equivalently fewer parameters. Multi-axis gated MLP is in the following equation (1) for complexity analysis:

%\begin{math}

\[\omega(MAB) = d2HWC (Global gMLP) + b2HWC (Local gMLP) + 10 HWC2 (Dense layers),          (1) \]

%\[\sqrt{x^2+1}\] 
%\end{math}

Losses accumulate throughout stages and sizes during MAXIM’s end-to-end training in below the following equation (2): 

L = s=1Sn=1NLchar(Rs,n,Tn) + Lfreq(Rs,n,Tn)                       (2)

Where Tn denotes multi-scale images, Lchar  and Lfreq are the Charbonnier loss and the frequency reconstruction loss.

\begin{figure}
    \centering
    \includegraphics[width=\linewidth]{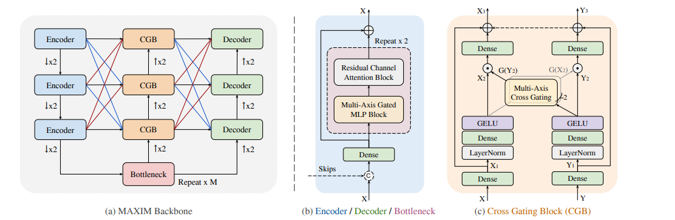}
    \caption{MAXIM architecture of ( a. Backbone, b. Encoder/ Decoder/ Bottleneck, c. Cross Gating Block) ref.4}
    \label{fig10}
\end{figure}

In a different work, Tolle et. al. suggested a novel Bayesian approach to deep imaging by utilizing mean field variational inference (MFVI)~\cite{7}. It has been mainly used in denoising, super-resolution, and inpainting for image processing. This approach permits for uncertainty quantification on a per-pixel layer distributed on the neural weights and omits the necessity of early stopping. Using Bayesian optimization with Gaussian Process Regression (GRP) optimizes the parameters for reconstruction accuracy. We demonstrate that a poorly chosen prior results in inferior accuracy and validation and that optimizing the weight pre-value for each cognitive system is adequate. We have shown below the mathematical notion of the MFVI DIP (figure 3).

\begin{figure}
    \centering
    \includegraphics[width=\linewidth]{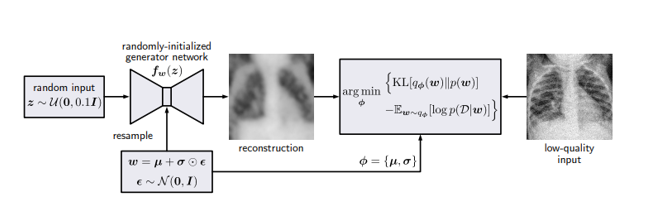}
    \caption{Demonstration of the mathematical notion abaft MFVI DIP \cite{5}}
    \label{fig1}
\end{figure}

Since the image processing task is very crucial, Cheng and his research team build a pre-trained model which is named image processing transformer (IPT). It has mainly been the representation of the transformer and its variant architectures. In this paper, the IPT model has worked on low-level computer vision tasks and developed a new pre-trained model~\cite{6}.In essence, it is indicated by the popular ImageNet benchmark datasets. This IPT architecture consists of heads, transformer encoder/decoder, and tails have been shown in Figure 5. Apparently, this new pre-trained model is set up well suited to various image processing tasks and therefore gets the desired score after several tuning parameters.

\begin{figure}
    \centering
    \includegraphics[width=\linewidth]{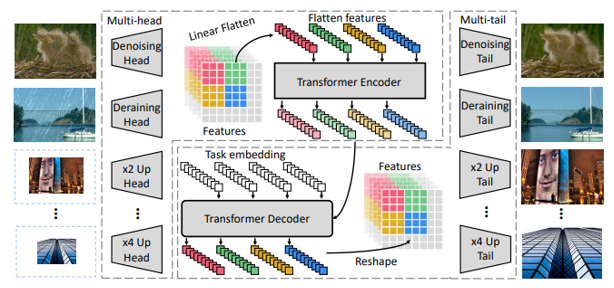}
    \caption{this image processing transformers suggested schematic (IPT). The IPT model includes an encoder and a decoder in addition to many heads and tails for various functions that share a common transformer block~\cite{8}.}
    \label{fig2}
\end{figure}

In another study, Singh did an empirical analysis that weighed the impact of image segmentation on skin lesion detection. Here, 10 deep-learning-based models to identify and classify the eight existing image segmentation methods. In this study, image enhancement and image morphology were used for picture processing for getting image classification performance and employed a double experimental design. The ResNet50 model has given the best performance with 91.9\% accuracy on the ISIC2017 dataset and compared to original photos with producing segmented images [9]. In the subject of the Internet of Medical Things (IoMT), one of the vital aspects is medical image processing. In recent times, deep learning algorithms have produced effective results on problems requiring the identification of medical images. As most of the time, we get some problems when we use typical deep learning algorithms only for causing little training data and domain mismatch. In the recent case which was Covid19, we could not detect covid19 by perusing computed tomography (CT) images. For this reason, Niu et. al. developed a well-known approach called distant domain transfer learning (DDTL)~\cite{10}. It also has privacy policies so that training data can not be easily accessed. In addition, It consists of two components: the reduced-size Unet segmentation model and the Distant Feature Fusion (DFF) classification model as shown in Figure 4. It has achieved better accuracy when we have tested unseen data using evaluation metrics.

\begin{figure}
    \centering
    \includegraphics[width=\linewidth]{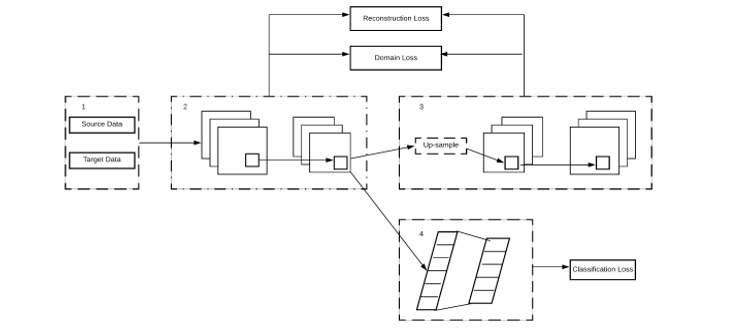}
    \caption{Distant Feature Extraction (DFF) architecture [8]}
    \label{fig3}
\end{figure}

Zaghl et. al. focused on the classification of melanoma skin cancer only using image processing techniques. So in this research, they are approached with four steps such as (i) enhancing algorithms, (ii) segmentation stage, (iii) feature extraction, and finally (iv) measured total dermoscopy value (TDV) for cancer classification. Therefore, day-by-day image processing
techniques are important for medical image analysis to get a proper result~\cite{11}. For both diagnosis and therapy, a variety of medical imaging techniques are employed. The most popular ones include MRI, X-ray, ultrasound, radionuclide, and optical. So Nagornov et. al suggests a system based on RNS for FPGA (field programmable gate array). For the purpose of processing 2D and 3D medical images, the discrete wavelet transform is one way to apply different fusion, denoising, and compression techniques. With the evolution of scanning technology and digital gadgets, medical imaging systems provide more accurate images~\cite{12,shah2019brain}. FPGA accelerators treat wavelet processing (WP) with scaled filter coefficients(SFC) and parallel computing to ameliorate the effects of top-notch 3D picture systems. This method enhances device performance by 2.89-3.59 times and raises the hardware resources by 1.18-3.29 times. In Figure 5 we have shown wavelet processing for medical images.

\begin{figure}
    \centering
    \includegraphics[width=\linewidth]{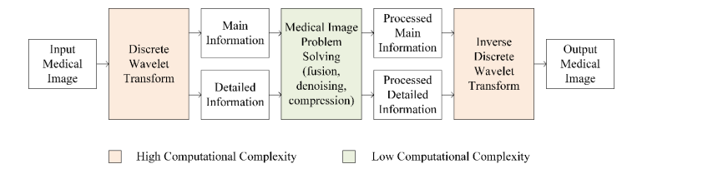}
    \caption{Wavelet processing for medical images. [10]}
    \label{fig4}
\end{figure}

Computerized diagnostic picture segmentation is crucial for facilitating and accelerating healthcare practices' screening and therapeutic processes. Therefore, Karimzadeh and his research team have made a novel model with morphology processing methods on CNN and PCA architecture. As a shape-based loss function had applied in lieu of  Binary-Cross-Entropy that is the Dice scores enhanced from 0.81$\pm$0.03 and 0.74$\pm$0.07 to 0.86$\pm$0.03 and 0.87$\pm$0.05 respectively for segmentation from MR and CT images. So the suggested PCA-based loss function was used and there were no outliers, patchy, or unrealistic metrics~\cite{13}.

% Please add the following required packages to your document preamble:
% \usepackage{booktabs}
% \usepackage{graphicx}
\begin{table}[htbp!]
\caption{Several models with various image processing methods for a better outcome on Medical Images}
\label{tab:my-table}
\resizebox{\textwidth}{!}{%
\begin{tabular}{@{}llllll@{}}
\toprule
\textbf{\begin{tabular}[c]{@{}l@{}}Author   \\ \& Year\end{tabular}} & \textbf{Dataset} & \textbf{\begin{tabular}[c]{@{}l@{}}Image \\ Processing \\ Techniques\end{tabular}} & \textbf{Model} & \textbf{Result} & \textbf{Conclusion} \\ \midrule
\begin{tabular}[c]{@{}l@{}}Tu et. al. \\ (2022)\end{tabular} & \begin{tabular}[c]{@{}l@{}}SIDD, DND,  \\ GoPro, HIDE, \\ REDS,\\  RealBlur-\\ RJ,Rain13k, \\ Raindrop,\\  RESIDE-   \\ (Indoor, \\ Outdoor), Five-\\ K, LOL\end{tabular} & \begin{tabular}[c]{@{}l@{}}Denoising, \\ Deblurring, \\ Deraining,   \\ Dehazing, \\ Enhancement\end{tabular} & \begin{tabular}[c]{@{}l@{}}MAXIM-3S, \\ MAXIM-2S\end{tabular} & \begin{tabular}[c]{@{}l@{}}PSNR \& SSIM, \\ this metrics are \\ given better \\ accuracy with\\ using MAXIM\\  model.\end{tabular} & \begin{tabular}[c]{@{}l@{}}Affordable \& \\ Productive for \\ image processing.\end{tabular} \\
\begin{tabular}[c]{@{}l@{}}Karimzadeh \\ et. al. \\ (2022)\end{tabular} & \begin{tabular}[c]{@{}l@{}}Decathlon \\ Medical Image\\ segmentation \\ challenge (MR), \\ SegTHOR   (CT)\end{tabular} & \begin{tabular}[c]{@{}l@{}}Image shape,\\  morphology\end{tabular} & CNN, PCA & \begin{tabular}[c]{@{}l@{}}Dice Scores \\ 0.81±0.03 and \\ 0.74±0.07 to \\ 0.86±0.03 and \\ 0.87±0.05 f\end{tabular} & \begin{tabular}[c]{@{}l@{}}Strengthen \\ realistic, reliable shape\\ estimations\end{tabular} \\
\begin{tabular}[c]{@{}l@{}}Nagaronov\\ et. al. (2022)\end{tabular} & \begin{tabular}[c]{@{}l@{}}Medical Image \\ ( tomographic)\end{tabular} & \begin{tabular}[c]{@{}l@{}}Fusion, \\ Denoising,\\ Compression\end{tabular} & \begin{tabular}[c]{@{}l@{}}Wavelet \\ processing \\ with SFC \\ and RNS\end{tabular} & \begin{tabular}[c]{@{}l@{}}Performance \\ 2.89-3.59 times.\end{tabular} & \begin{tabular}[c]{@{}l@{}}The use of SFC \\ and RNS\\ improved the \\ efficiency of \\ 3D medical\\ image WP \\ devices.\end{tabular} \\
Singh (2022) & \begin{tabular}[c]{@{}l@{}}ISIC2016, \\ ISIC2017, and \\ ISIC2018 \\ (skin-lesion),   \\ PH2\end{tabular} & \begin{tabular}[c]{@{}l@{}}Image \\ Enhancement \& \\ Morphological\end{tabular} & \begin{tabular}[c]{@{}l@{}}10 DL \\ Models\\ (VGG16,19, \\ ResNet50,\\ etc.)\end{tabular} & \begin{tabular}[c]{@{}l@{}}Best accuracy = \\ 91.2\% (Otsu’s \\ Binarization \\ ResNet50)\end{tabular} & \begin{tabular}[c]{@{}l@{}}Segmentation \\ for \\ classification\\ performance on \\ skin-lesion   \\ datasets.\end{tabular} \\
Niu (2021) & \begin{tabular}[c]{@{}l@{}}Catech-256, \\ Office-31, \\ Chest Xray, \\ Lung-CT, \\ Covid19-CT\end{tabular} & \begin{tabular}[c]{@{}l@{}}Feature \\ Extraction, Removal \\ noise, Image \\ Reconstruction\end{tabular} & \begin{tabular}[c]{@{}l@{}}Distant \\ Domain \\ Transfer \\ Learning \\ (DDTL)\end{tabular} & \begin{tabular}[c]{@{}l@{}}Acquired  96\% \\ accuracy\end{tabular} & \begin{tabular}[c]{@{}l@{}}Novel DDTL, \\ DFF, a \\ diagnostic \\ method for lung\\ CT images.\end{tabular} \\
\begin{tabular}[c]{@{}l@{}}Tolle \\ et. al. (2021)\end{tabular} & \begin{tabular}[c]{@{}l@{}}OCT scans, \\ Chest X-rays,\\ MRI Public \\ datasets, HAM10000 \\ datasets\end{tabular} & \begin{tabular}[c]{@{}l@{}}Denoising, \\ Super-\\ Resolution, \\ Inpainting\end{tabular} & \begin{tabular}[c]{@{}l@{}}Bayesian \\ approach \\ with MFVI\end{tabular} & \begin{tabular}[c]{@{}l@{}}Denoising = \\ 0.05, \\ Super-Resolution = 0.1,\\  Inpainting = 0.36\end{tabular} & \begin{tabular}[c]{@{}l@{}}MFVI approach \\ to DIP and \\ optimized the \\ weight by \\ Bayesian   \\ Optimization.\end{tabular} \\
\begin{tabular}[c]{@{}l@{}}Chen \\ et. al. \\ (2021)\end{tabular} & \begin{tabular}[c]{@{}l@{}}ImageNet \\ Datasets\end{tabular} & \begin{tabular}[c]{@{}l@{}}Super-\\ Resolution, \\ Denoising, \\ Deraining,\end{tabular} & \begin{tabular}[c]{@{}l@{}}Image \\ Processing \\ Transformer (IPT)\end{tabular} & \begin{tabular}[c]{@{}l@{}}IPT given the \\ best result on \\ different datasets.\end{tabular} & \begin{tabular}[c]{@{}l@{}}Designed with \\ multi-heads, \\ multi-tails, and \\ extending to \\ more   \\ inpainting, \\ dehazing.\end{tabular} \\
\begin{tabular}[c]{@{}l@{}}Zghal \\ et. al. \\ (2020)\end{tabular} & PH2 Database & \begin{tabular}[c]{@{}l@{}}Image \\ acquisition, \\ filtering \& \\ contrast \\ enhancement, \\ Feature extract\end{tabular} & \begin{tabular}[c]{@{}l@{}}Asymmetry \\ Border Color \\ Diameter, \\ SVM, TDV\end{tabular} & \begin{tabular}[c]{@{}l@{}}Accuracy  90\%,\\    \\ Specificity 92\%, \\ Sensitivity 87\%.\end{tabular} & \begin{tabular}[c]{@{}l@{}}Automatically \\ detects \\ pigmented skin \\ lesions and \\ diagnoses   \\ melanoma.\end{tabular} \\ \bottomrule
\end{tabular}%
}
\end{table}

\section{Feature Engineering}

With the process of invention, innovation and diffusion, scientific advancement has reached a level which is inconceivable. However, feature engineering, one of the advancements, has been at the center of attention for a while. Complications arise when there is a huge amount of data available to use for Artificial intelligence and machine learning techniques and the necessity for feature engineering appears. Feature engineering is a multistep process consisting of the creation, extraction, and selection of variables that are most accurate. This process converts raw data into features which makes any case easy to analyze.

\begin{figure}
    \centering
    \includegraphics[width=\linewidth]{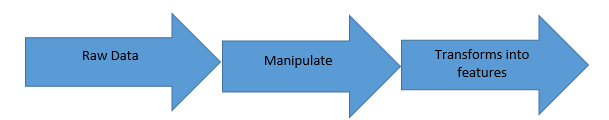}
    \caption{Feature Engineering basic}
    \label{fig5}
\end{figure}

While defining features, for instance, when we work with observations in tabular data, each observation has attributes, which depict something meaningful about the observation referred to as features. Feature engineering art varies among data scientists. Steps to perform feature engineering are given below~\cite{14}:

Raw data collection:  Collecting data from different sources which may encompass unstructured, structured, textual data.

Data Processing: The process consists of raw data manipulation and unification from different sources, which may involve data amplification, data augmentation, fusion, ingestion, stochastic simulation, and sampling error.

Feature Creation:
Visualize and plot data if something is not working with adding up data then the data is filtered to create features to be used for modeling the data. It involves domain knowledge, instinct, and a lengthy process of trial and error. The human attention involved in administering this process significantly shapes the cost of model generation \cite{al2022brain,islam2021brain}.

Feature selection: Algorithms are responsible to analyze and judging features to determine which features to take into account or which features are redundant or irrelevant and should be removed.

Modeling: To evaluate the quality of selected features, models are created which may utilize the execution of learning algorithms, for instance, cross-validation, wrapper models etc.

Benchmark: The reduction rate of error and improvement of the model's prognostication and accuracy standard is set where all variables are compared. This is the stage where data scientists with domain expertise do experiments and testing for benchmarking.

\begin{figure}
    \centering
    \includegraphics[width=\linewidth]{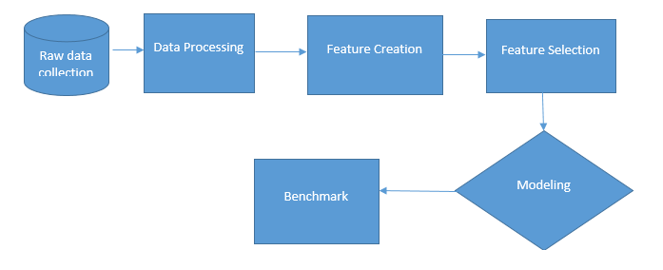}
    \caption{Steps for feature engineering}
    \label{fig6}
\end{figure}

The acceptance or declination of a predictive model is determined by feature engineering. In machine learning, feature engineering imposes data for creating new variables to enhance model accuracy. To provide a prediction, different machine learning models, for instance, decision trees, random forests, neural networks, and gradient boosting machines accept a feature vector. From the provided feature set, new features are engineered. Primarily, this process is manual and will be different for different kinds. Feature vector plays a crucial role in machine learning as most machine learning performance is dependent on it [15]. Although, a number of automated feature tools are available, for instance, Deep feature synthesis (relational and temporal data), Precise handling of time (keeping data safe from common label leakage problems, prediction time row by row can be specified), reusable feature primitive (It is possible to build and share own custom primitives to reuse on any dataset)[16]. For better analysis, feature engineering is seen as a generalization of mathematical optimization \cite{nasim2021prominence}.

\section{Machine Learning}

Systems can learn and develop automatically based on experience allowing a technique called machine learning. Without specialist programming, machine learning is capable of doing tasks. Machine learning refers to the process of developing computer programs that can examine data and decide for themselves via the use of a set of algorithms what should be conducted with that information~\cite{ali2021alzheimer}.
The most prevalent algorithm is supervised machine learning. A supervised machine learning method uses prior information to interpret new input. The system is reading a fresh but comparable data set, and supervised algorithms use examples from earlier, related training data sets to identify flaws within them to predict future problems based on those instances. Often, training data sets are initially determined by humans, and then the system is trained to recognize patterns associated with each pertinent training data set. After that, the system compares the recently obtained data sets to the training data sets. The algorithm can identify and anticipate certain problems after having access to sufficient training data sets for comparison. This type of machine learning can uncover defects that improve the relevant model by further comparing its selected response to the intended actions that were corrected. Unsupervised machine learning methodologies make use of unlabeled "raw" data.
This raw data has no known flaws, therefore it only attempts to infer an action based on concealed unlabeled flaws inside uncategorized data. Because the system lacks a predefined failure mode pattern or any suggestions for possible answers, it cannot determine the appropriate course of action. Nonetheless, the system examines the information and makes an attempt to inferences based on irregularities or flaws that have been tagged in any learning data set.  Semi-supervised machine learning techniques integrate supervised and unsupervised learning by utilizing both labeled and unlabeled data sets for training. These sets of data frequently include substantial volumes of unlabeled, uncategorized information. After that, the system aggregates smaller, explicitly stated data points on labeled and classified fault patterns. Algorithms for reinforcement machine learning employ trial-and-error behavior based on a data set, and they award themselves whenever the correct action is taken.
As a consequence, the system uses a reinforcing technique in its decision-making mechanism. Based on clear reward feedback, it can quickly and automatically determine the optimum course of action within a specific data set. In order to boost the system's performance, this method immediately employs reinforcement signals. The following Figure 1 is a machine learning process in which we maintain these steps.

\begin{figure}
    \centering
    \includegraphics[width=\linewidth]{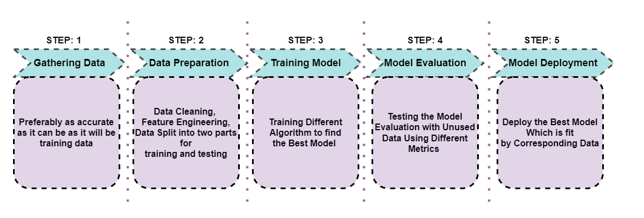}
    \caption{Machine Learning Process}
    \label{fig7}
\end{figure}

Machine learning is a subfield of artificial intelligence that focuses on developing algorithms and models that can learn from data and make predictions or take actions without being explicitly programmed. To perform machine learning, an algorithm is trained on a dataset. The algorithm can then make predictions or decisions based on the patterns it has learned from the data. 

To understand it more, we can think of a story. Dr. Sarah was a radiologist at a busy hospital, responsible for interpreting medical images and diagnosing patients based on the images. She spent long hours poring over X-rays, CT scans, and MRIs, trying to identify abnormalities and make accurate diagnoses.

One day, Dr. Sarah was introduced to a new machine-learning model developed to assist with interpreting medical images. At first, she was sceptical, but as she started using the model, she was amazed by its accuracy and speed.

The machine learning model had been trained on thousands of medical images and the corresponding diagnoses made by expert radiologists. As a result, it could identify patterns and abnormalities in the images that were not immediately obvious to the human eye.

Dr. Sarah started using the machine learning model to assist with her diagnoses, which helped her identify issues more quickly and accurately than before. She could also see more patients in a day, thanks to the time the machine learning model saved her.

\subsection{How machine learning model learn}

A machine learning model learns by being trained on a dataset of labeled examples. The process of training a machine learning model involves providing the model with a large number of examples that have been labeled with the correct output (also known as the ground truth). The model uses these labeled examples to learn how to map input data to the correct output.

\subsection{Types of machine learning}

There are three main types of machine learning:

\subsubsection{Supervised learning.}
 In supervised learning, the algorithm is trained on labeled data, where the correct output is provided for each input. Supervised learning aims to make predictions or decisions based on the patterns learned from the data. Examples of supervised learning include predicting whether a customer will churn based on their past behavior or identifying the sentiment of a tweet as positive or negative.

\subsubsection{Unsupervised learning.}
 In unsupervised learning, the algorithm is not provided with labeled data and must discover patterns independently. Unsupervised learning aims to identify patterns or relationships in the data. Examples of unsupervised learning include clustering data points into groups based on their similarities or identifying fraudulent transactions based on unusual patterns in the data.

 \subsubsection{Reinforcement learning.}
 In reinforcement learning, the algorithm learns through trial and error, receiving rewards or penalties for certain actions. The goal of reinforcement learning is to learn the best action to take in a given situation to maximize a reward. Examples of reinforcement learning include training a robot to navigate through a maze by rewarding it for reaching the end and penalizing it for making incorrect turns or training a self-driving car to make decisions based on the environment and traffic conditions.

 \subsection{Limitations of machine learning}
 
Healthcare is one of the many areas that machine learning has the potential to disrupt. When creating and using these systems, developers and implementers should be aware of several machine-learning restrictions.

The quality and quantity of data available for training is one restriction. Machine learning algorithms need high-quality data to learn and produce precise predictions. Due to the complexity of the data, the need for more consistency in data collection, and the difficulty in getting significant quantities of patient data that are representative of the population, this can be problematic in the healthcare industry.

The models' interpretability is yet another drawback. Although machine learning models may produce precise predictions, it can be challenging to comprehend how they do so. This might make it challenging to pinpoint the mistakes' root causes and modify the model.

Another restriction is that machine learning algorithms are frequently created for certain tasks and may need to generalize better to other activities. This might make applying a model trained for one job to another challenging.

Furthermore, the quality of machine learning models depends on the data they are trained on. If the data are skewed, the model will pick up on that bias and spread it through its forecasts. This is a prevalent issue in the healthcare industry, where inequalities exist and a lack of diversity in data collecting and representation.

\section{Deep Learning}

Deep learning is a subfield of machine learning that involves using artificial neural networks with many layers (hence the term "deep") to learn and make decisions based on data. Deep learning has achieved remarkable success in many applications, including image and speech recognition, natural language processing, and self-driving cars.

One of the key benefits of deep learning is its ability to learn and make decisions based on raw, unstructured data, such as images or text. Traditional machine learning techniques often require that the data be manually extracted and processed, which can be time-consuming and error-prone. On the other hand, deep learning can learn directly from the raw data, making it more efficient and accurate.

Deep learning is implemented using artificial neural networks inspired by how the human brain works. An artificial neural network consists of layers of interconnected "neurons," each receiving input and producing an output based on weights and biases. These weights and biases are adjusted during training to learn the relationships between the input data and the desired output~\cite{ahsan2023deep}.

Deep learning has achieved impressive results in various applications and is expected to play a significant role in the future of machine learning and artificial intelligence. However, it is important to note that deep learning is not a magic bullet and can still be affected by issues such as bias in the training data and overfitting. Noisy labels \cite{karim2022unicon} can compromise generalization ability of medical imaging models

\begin{figure}
    \centering
    \includegraphics[width=\linewidth]{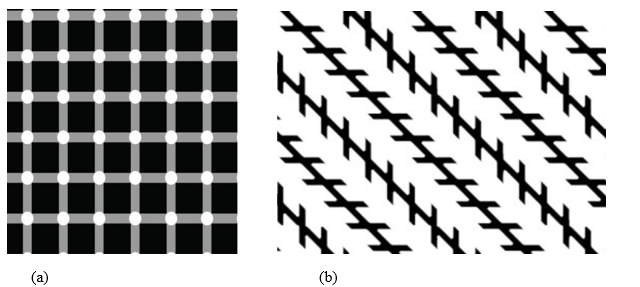}
    \caption{Some optical illusions tell us about the visual system. (a) Looking at this deceiving picture, the brain thinks there’s a black dot inside each white circle until someone focuses on each individual white circle and in the end realizes that it was never at all. (b) These long diagonal lines are parallel. They sure don’t look it, but they are! Removing the smaller “stitch”-like lines shows the truth about this optical illusion..}
    \label{fig8}
\end{figure}

In the 1970s, computer vision first started and was viewed as the visual perception of an aspiring agenda to mimic human intelligence and endow robots with intelligent behavior. Later in the 1980s, priority was given to the sophistication of mathematical techniques to perform quantitative image and scene analysis. Stepping ahead to the 2000s, data-driven and learning approaches as core constituents of vision is being embraced by this decade~\cite{biswas2022mri}.

\begin{figure}
    \centering
    \includegraphics[width=\linewidth]{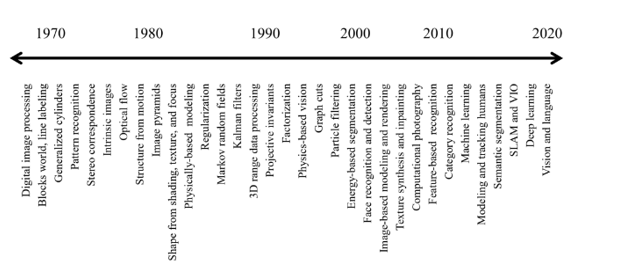}
    \caption{A rough record of some of the dynamic topics of research in computer vision}
    \label{fig9}
\end{figure}

\subsection{How the deep learning model learns}

Artificial neural networks with numerous layers, or "deep" neural networks, are the foundation of the powerful machine learning method. Instead of depending on fixed characteristics, these models are built to autonomously learn features and representations from the input. This makes deep learning models especially suitable for applications like audio and picture identification, natural language processing, and other areas where conventional machine learning models may fail. A standard machine learning model's learning process is similar to a deep learning model in that it is trained on one dataset before being tested on another to assess its performance. The way deep learning models learn, as opposed to conventional machine learning models, differs in a few significant ways. Deep learning models provide several key advantages over traditional models, including learning hierarchical data representations. As the model advances through the network's layers, it can understand increasingly abstract terms of the input. When classifying images, for instance, the network's first layers can pick up on essential elements like borders.
In contrast, its deeper layers might pick up on more intricate features like forms and objects. Another distinction between deep learning and regular machine learning models is that the latter often require structured data, while the former may learn from unstructured data like text and pictures. Deep learning models are particularly well-suited for tasks like speech and picture recognition. Furthermore, unlike typical machine learning models, which could have trouble with massive datasets, deep learning models can learn from enormous volumes of data. More advanced than typical machine learning models, deep learning models can learn from the data without supervision. Finally, deep learning models are more generalizable to new data than conventional machine learning models. This is so that deep learning models can acquire more abstract data representations that are less reliant on the particulars of the training set. In general, deep learning is a powerful method that can enhance the effectiveness, accuracy, and dependability of services within the medical business and offer new chances for the area of medical imaging informatics to create new paths toward precision medicine.

\subsection{Different types of deep learning models}

There are several different types of deep learning models, including:

\begin{enumerate}
  \item Convolutional neural networks (CNNs): CNNs are used for image and video analysis and are particularly well-suited for tasks such as image classification, object detection, and image segmentation\cite{tonmoy2019brain}. They are called "convolutional" because they use convolutional layers, which are used to extract features from the input data.
  \item Recurrent neural networks (RNNs): RNNs are used for tasks that involve sequential data, such as natural language processing, speech recognition, and time series analysis. They are called "recurrent" because they have loops that allow them to process data over time.
  \item Long short-term memory (LSTM) networks: It is a type of RNN that are particularly effective at learning long-term dependencies in sequential data. They have been used for tasks such as language translation, language modeling, and speech recognition.
  \item Autoencoders: Autoencoders are a type of neural network used for dimensionality reduction and feature learning. They consist of an encoder that maps the input data to a lower-dimensional representation and a decoder that maps the lower-dimensional representation back to the original input space.
  \item Generative adversarial networks (GANs): GANs is a neural network that generates new data similar to a training dataset. They consist of two networks: a generator network that generates new data and a discriminator network that distinguishes the generated data from the actual data.
  \item Transfer learning: Transfer learning is a technique in which a deep learning model trained on one task is fine-tuned for a different task. This can be useful when more data is needed to train a model from scratch.
\end{enumerate}

These are just a few examples of the many deep-learning models available. The appropriate model for a particular task will depend on the type and complexity of the data, as well as the specific goals of the model.

\subsection{Limitations of deep learning}

Deep learning has made strides in several fields, including self-driving cars, audio and picture identification, and natural language processing. It uses neural network models layers of linked nodes—to automatically learn from a lot of data and gradually improve performance. However, it's crucial to remember that deep learning is not a universally applicable solution and has several drawbacks that must be considered before using it for a particular task. One of the significant drawbacks is the need for a lot of labelled data, which can be difficult when there isn't much of it or when it has to be accurately labelled and indicative of real-world settings. This may be especially troublesome in industries like healthcare, where patient data is frequently private and hard to get.

The inability to analyze deep learning models is another drawback. It may be challenging to comprehend the reasoning behind a particular choice or forecast since the models tend to learn complicated correlations between the input and output data. The inability to understand the results makes finding and fixing model flaws challenging. Deep learning models may also be prejudiced if the training data contains any biases. For instance, the model may be biased toward identifying white individuals if the training data contains more photographs of white people than images of people of other races.

Another restriction that might result in worse performance on fresh, untried data is overfitting, which occurs when the model is too complicated and has learned too much from the training data. This is very troublesome in real-world applications where the data is not the same as the training data. Finally, deep learning models need a lot of computer power to train, which might be difficult if there aren't enough. This can be especially troublesome for small and medium-sized businesses or organizations, which might need access to the same resources as big companies.

\section{Importance of data in machine learning and deep learning}

Data is the foundation of machine learning and deep learning algorithms. The algorithms are trained on data, and the quality and quantity of the data used to train the model significantly impact its performance. To build an accurate and effective model, it is essential to have high-quality data relevant to the problem being solved.

The importance of data in machine learning and deep learning can be understood by considering its role. First, the data is used to train the model, which involves feeding it to the algorithm and adjusting its internal parameters to minimize the error between the model's predictions and the true outcomes. The quality and quantity of the data used to train the model directly impact its ability to learn and make accurate predictions.

Second, the data is used to evaluate the model's performance. Once the model has been trained, it is important to evaluate its performance on a separate test set. This allows the model's performance to be assessed and indicates how well the model is likely to perform on new, unseen data.

Finally, the data is used to fine-tune the model. Suppose the model's performance could be more satisfactory. Several techniques can be used to fine-tune it, such as adjusting the model's hyperparameters or adding additional layers or units to the model. The data used to fine-tune the model can be used to identify areas where the model is performing poorly and help to improve its overall performance.

\section{Recent advancements in computer vision}

Computer vision is a field that deals with how computers can be made to understand and interpret the visual world. It has made tremendous progress in recent years, thanks to advances in machine learning and hardware capabilities.
There have been many significant advancements in the field of computer vision in recent years. Some of the most notable include:

\subsubsection{Deep learning} The use of deep neural networks has significantly improved the performance of computer vision systems in tasks such as object recognition, object detection, and image segmentation.

\subsubsection{Transfer learning} Pre-trained deep learning models can be fine-tuned for specific tasks, allowing for more efficient training and improved performance on small datasets.

\subsubsection{Generative adversarial networks (GANs)} GANs have been used to generate realistic images and applied to tasks such as image-to-image translation and image super-resolution.

\subsubsection{Computer vision in robotics} Advances in computer vision have enabled the development of robots that can navigate and interact with their environment using visual information.

\subsubsection{Augmented reality (AR)} Computer vision techniques are used in AR systems to detect and track objects in the real world, allowing for the overlay of digital content onto the physical world.
\subsubsection{Video analysis} Deep learning models have been applied to action recognition, scene understanding, and video content summarization tasks.
\subsubsection{Medical image analysis} Computer vision techniques are being used to analyze medical images, such as CT scans and X-rays, allowing for the automatic detection of abnormalities and the development of assistive tools for physicians.
\section{Conclusion and Future Direction}

For over three decades Medical imaging informatics has been driving clinical research, translation and practice. Already deep in the big medical data era, imaging data availability is only expected to grow, complemented by massive amounts of associated data-rich EMR/ HER and physiological data, climbing to orders of magnitude higher than what is available today. As such, the research community is struggling to harness the full potential of the wealth of data that are now available at the individual patient level underpinning precision medicine. Keeping up with storage, sharing, and processing while preserving privacy and anonymity has pushed boundaries in traditional means of doing research. Imaging investigators often have issues with managing data, indexing, investigating and query digital pathology data. One of the abstract challenges is how to manage relatively comprehensive, multi-layered data sets that will continue to amplify over time since it is unjustifiable to intensively compare the query data with each sample in a high dimensional database due to practical storage and processing constraints [33]. How to thoroughly scrutinize the characteristics of data commencing from multiple approaches, is a riddle. Data analytics approaches, such as machine learning and statistical analysis, can be used to automatically identify and analyze important features in large data sets, including anatomical areas of interest and physiological phenomena. This can help researchers and scientists to better understand the underlying physiology and pathophysiology of different tissues and regions in the body. It would be impossible to gain insights and make discoveries without using these approaches. Particularly in the field of artificial intelligence and machine learning, deep learning methods are currently being used in a lot of research endeavours. Although challenges exist, researchers are actively working to address challenges. Some of the key areas of focus include developing explainable AI methods which can help to better understand how these systems make decisions, as well as leveraging advanced techniques such as 3D reconstruction and visualization to improve the performance of these systems. In conclusion, medical imaging informatics advances are anticipated to improve the quality of care levels witnessed today, once innovative solutions along the lines of selected research endeavors presented in this study are adopted in clinical practice and thus potentially transforming precision medicine.

\bibliographystyle{unsrt}
\bibliography{main.bib}

\begin{thebibliography}{10}

\bibitem{1}
Casimir~A Kulikowski.
\newblock Medical imaging informatics: challenges of definition and
  integration, 1997.

\bibitem{2}
Alex~AT Bui and Ricky~K Taira.
\newblock {\em Medical imaging informatics}.
\newblock Springer Science \& Business Media, 2009.

\bibitem{3}
Andreas~S Panayides, Amir Amini, Nenad~D Filipovic, Ashish Sharma, Sotirios~A
  Tsaftaris, Alistair Young, David Foran, Nhan Do, Spyretta Golemati, Tahsin
  Kurc, et~al.
\newblock Ai in medical imaging informatics: current challenges and future
  directions.
\newblock {\em IEEE Journal of Biomedical and Health Informatics},
  24(7):1837--1857, 2020.

\bibitem{4}
William Hsu, Mia~K Markey, and May~D Wang.
\newblock Biomedical imaging informatics in the era of precision medicine:
  progress, challenges, and opportunities.
\newblock {\em Journal of the American Medical Informatics Association},
  20(6):1010--1013, 2013.

\bibitem{5}
Angela Giardino, Supriya Gupta, Emmi Olson, Karla Sepulveda, Leon Lenchik, Jana
  Ivanidze, Rebecca Rakow-Penner, Midhir~J Patel, Rathan~M Subramaniam, and
  Dhakshinamoorthy Ganeshan.
\newblock Role of imaging in the era of precision medicine.
\newblock {\em Academic radiology}, 24(5):639--649, 2017.

\bibitem{6}
C~Chennubhotla, LP~Clarke, A~Fedorov, D~Foran, G~Harris, E~Helton, R~Nordstrom,
  F~Prior, D~Rubin, JH~Saltz, et~al.
\newblock An assessment of imaging informatics for precision medicine in
  cancer.
\newblock {\em Yearbook of medical informatics}, 26(01):110--119, 2017.

\bibitem{9}
Jack Sklansky.
\newblock Image segmentation and feature extraction.
\newblock {\em IEEE Transactions on Systems, Man, and Cybernetics},
  8(4):237--247, 1978.

\bibitem{10}
Robert~J Gillies, Paul~E Kinahan, and Hedvig Hricak.
\newblock Radiomics: images are more than pictures, they are data.
\newblock {\em Radiology}, 278(2):563, 2016.

\bibitem{11}
Hugo~JWL Aerts, Emmanuel~Rios Velazquez, Ralph~TH Leijenaar, Chintan Parmar,
  Patrick Grossmann, Sara Carvalho, Johan Bussink, Ren{\'e} Monshouwer,
  Benjamin Haibe-Kains, Derek Rietveld, et~al.
\newblock Decoding tumour phenotype by noninvasive imaging using a quantitative
  radiomics approach.
\newblock {\em Nature communications}, 5(1):1--9, 2014.

\bibitem{12}
Nilesh~Bhaskarrao Bahadure, Arun~Kumar Ray, and Har~Pal Thethi.
\newblock Image analysis for mri based brain tumor detection and feature
  extraction using biologically inspired bwt and svm.
\newblock {\em International journal of biomedical imaging}, 2017, 2017.

\bibitem{13}
Justin Ker, Lipo Wang, Jai Rao, and Tchoyoson Lim.
\newblock Deep learning applications in medical image analysis.
\newblock {\em Ieee Access}, 6:9375--9389, 2017.

\bibitem{14}
Fangzhou Liao, Ming Liang, Zhe Li, Xiaolin Hu, and Sen Song.
\newblock Evaluate the malignancy of pulmonary nodules using the 3-d deep leaky
  noisy-or network.
\newblock {\em IEEE transactions on neural networks and learning systems},
  30(11):3484--3495, 2019.

\bibitem{biswas2022mri}
Angona Biswas and Md~Saiful Islam.
\newblock Mri brain tumor classification technique using fuzzy c-means
  clustering and artificial neural network.
\newblock In {\em International Conference on Artificial Intelligence for Smart
  Community: AISC 2020, 17--18 December, Universiti Teknologi Petronas,
  Malaysia}, pages 1005--1012. Springer, 2022.

\bibitem{15}
S-CB Lo, S-LA Lou, Jyh-Shyan Lin, Matthew~T Freedman, Minze~V Chien, and
  Seong~Ki Mun.
\newblock Artificial convolution neural network techniques and applications for
  lung nodule detection.
\newblock {\em IEEE transactions on medical imaging}, 14(4):711--718, 1995.

\bibitem{hossainbrain}
Tonmoy Hossain, Fairuz~Shadmani Shishir, Mohsena Ashraf, MD~Abdullah
  Al~Nasim4\&, and Faisal~Muhammad Shah.
\newblock Brain tumor detection using convolutional neural network.

\bibitem{7}
Jerrold~T Bushberg and John~M Boone.
\newblock {\em The essential physics of medical imaging}.
\newblock Lippincott Williams \& Wilkins, 2011.

\bibitem{8}
Lifeng Yu, Xin Liu, Shuai Leng, James~M Kofler, Juan~C Ramirez-Giraldo,
  Mingliang Qu, Jodie Christner, Joel~G Fletcher, and Cynthia~H McCollough.
\newblock Radiation dose reduction in computed tomography: techniques and
  future perspective.
\newblock {\em Imaging in medicine}, 1(1):65, 2009.

\bibitem{nasim2021prominence}
MD~Nasim, Aditi Dhali, Faria Afrin, Noshin~Tasnim Zaman, and Nazmul Karim.
\newblock The prominence of artificial intelligence in covid-19.
\newblock {\em arXiv preprint arXiv:2111.09537}, 2021.

\bibitem{shah2019brain}
Faisal~Muhammad Shah, Tonmoy Hossain, Mohsena Ashraf, Fairuz~Shadmani Shishir,
  MD~Abdullah Al~Nasim, and Md~Hasanul Kabir.
\newblock Brain tumor segmentation techniques on medical images-a review.

\bibitem{al2022brain}
Md~Abdullah Al~Nasim, Abdullah Al~Munem, Maksuda Islam, Md~Aminul~Haque Palash,
  Md~Mahim~Anjum Haque, and Faisal~Muhammad Shah.
\newblock Brain tumor segmentation using enhanced u-net model with empirical
  analysis.
\newblock In {\em 2022 25th International Conference on Computer and
  Information Technology (ICCIT)}, pages 1027--1032. IEEE, 2022.

\bibitem{islam2021brain}
Md~Khairul Islam, Md~Shahin Ali, Md~Sipon Miah, Md~Mahbubur Rahman,
  Md~Shahariar Alam, and Mohammad~Amzad Hossain.
\newblock Brain tumor detection in mr image using superpixels, principal
  component analysis and template based k-means clustering algorithm.
\newblock {\em Machine Learning with Applications}, 5:100044, 2021.

\bibitem{ali2021alzheimer}
Md~Shahin Ali, Md~Khairul Islam, Jahurul Haque, A~Arjan Das, DS~Duranta, and
  Md~Ariful Islam.
\newblock Alzheimer’s disease detection using m-random forest algorithm with
  optimum features extraction.
\newblock In {\em 2021 1st International Conference on Artificial Intelligence
  and Data Analytics (CAIDA)}, pages 1--6. IEEE, 2021.

\bibitem{ahsan2023deep}
Md~Manjurul Ahsan, Muhammad~Ramiz Uddin, Md~Shahin Ali, Md~Khairul Islam,
  Mithila Farjana, Ahmed~Nazmus Sakib, Khondhaker Al~Momin, and Shahana~Akter
  Luna.
\newblock Deep transfer learning approaches for monkeypox disease diagnosis.
\newblock {\em Expert Systems with Applications}, page 119483, 2023.

\bibitem{karim2022unicon}
Nazmul Karim, Mamshad~Nayeem Rizve, Nazanin Rahnavard, Ajmal Mian, and Mubarak
  Shah.
\newblock Unicon: Combating label noise through uniform selection and
  contrastive learning.
\newblock In {\em Proceedings of the IEEE/CVF Conference on Computer Vision and
  Pattern Recognition}, pages 9676--9686, 2022.

\bibitem{tonmoy2019brain}
Hossain Tonmoy, Shishir~Fairuz Shadmani, Ashraf Mohsena, MD~Al~Nasim Abdullah,
  and Muhammad~Shah Faisal.
\newblock Brain tumor detection using convolutional neural network.
\newblock In {\em 2019 1st International Conference on Advances in Science,
  Engineering and Robotics Technology (ICASERT), IEEE}, pages 1--6, 2019.

\end{thebibliography}

\end{document}